\documentclass[11pt]{article}

\usepackage[margin=1in]{geometry}
\usepackage[T1]{fontenc}
\usepackage[utf8]{inputenc}
\usepackage{lmodern}
\usepackage{setspace}
\usepackage{hyperref}
\usepackage{booktabs}
\usepackage{tabularx}
\usepackage{graphicx}
\usepackage{amssymb}

\title{Beyond Compliance: A Proposed Framework for Ethical Governance of Student Data in Learning Analytics}
\author{Sahana Varadaraju \quad Bharathwaj Vijayakumar \\[0.3em] Rowan University (United States)}
\date{}

\begin{document}
\maketitle

\begin{center}
\small
\textbf{Author manuscript prepared for arXiv.} This version contains minor revisions, including additional discussion, implementation guidance, comparison tables, and updated references relative to the published conference proceedings. The version of record is published in the EDULEARN26 Proceedings (IATED, 2026).\\[0.4em]
\textbf{Citation:} S. Varadaraju and B. Vijayakumar, \textit{Beyond Compliance: A Proposed Framework for Ethical Governance of Student Data in Learning Analytics}, EDULEARN26 Proceedings, IATED, 2026. DOI: \href{https://doi.org/10.21125/edulearn.2026.1927}{10.21125/edulearn.2026.1927}.
\end{center}

\begin{abstract}
The rapid growth of learning analytics (LA) in higher education has expanded institutional 
capacity to monitor engagement, predict academic difficulty, and target support using student data. 
While these practices offer important educational benefits, 
governance has often remained compliance-first, centered on meeting baseline legal requirements 
such as the Family Educational Rights and Privacy Act (FERPA) and the General Data Protection Regulation (GDPR). 
Legal compliance is necessary, but it does not by itself resolve questions of fairness, student agency, accountability, 
or educational purpose. This paper proposes the LEAGUE framework, a six-pillar model for ethical governance of student data in LA: Lawfulness, Equity, Agency, Governance, Utility, and Ethics by Design. 
The framework is developed through a conceptual synthesis of scholarship in learning analytics, educational data mining, data ethics, educational policy, 
value-sensitive design, and capability-oriented approaches to educational justice. 
Its practical value is demonstrated through an illustrative early alert case that shows how institutions can review learning analytics in a more transparent and educationally meaningful way.

\vspace{0.5em}
\end{abstract}

\section{Introduction}
Learning analytics has become a central part of higher education's digital environment. As universities rely more heavily on learning management systems, advising platforms, student information systems, and predictive tools, they generate large amounts of student data that can be used to monitor participation, estimate academic risk, and guide intervention \cite{long2011fog,macfadyen2010mining}. These systems are usually introduced in the name of student success. Institutions hope to identify disengagement earlier, coordinate support more efficiently, and make teaching and advising more responsive.

The challenge is that the ethical governance of student data has not developed at the same pace as the technologies used to collect and interpret them. In many institutions, the first question asked about a new analytics initiative is whether it is legally permissible. That question matters, but it is too narrow. A system can satisfy institutional policy and still be confusing to students, insensitive to context, weakly justified, or unfair in its effects.

This issue is especially important in learning analytics because student data are never self-explanatory. Login counts, time on platform, missed submissions, and attendance patterns do not speak for themselves. They must be interpreted, and those interpretations shape how students are classified and how support decisions are made \cite{slade2013ethical,knox2017data}. A student may appear inactive for reasons that have little to do with effort or ability. A risk score may create a useful prompt for outreach, but it may also conceal bias, encourage overconfidence, or reduce a student to a simplified label.

Existing laws and guidance provide an important baseline. FERPA and GDPR address privacy, records, disclosure, and data rights \cite{ferpa,eu2016gdpr}. Sector guidance such as Jisc's code of practice and Drachsler and Greller's DELICATE checklist for trusted learning analytics has also helped institutions think more carefully about responsible learning analytics \cite{jisc2023code,drachsler2016delicate}. Yet these approaches often focus on specific parts of the problem. What is still needed is a practical governance model that brings together legality, fairness, student agency, accountability, educational value, and design choices within one institutional framework.

This paper proposes such a model. The LEAGUE framework organizes ethical review around six pillars: Lawfulness, Equity, Agency, Governance, Utility, and Ethics by Design. The framework is intended as a practical tool for institutional review, and its usefulness is illustrated through a detailed early alert case.

\noindent\textbf{Contributions.} This paper makes four contributions. It (1) proposes the LEAGUE ethical governance framework; (2) integrates six governance dimensions into a unified institutional review model; (3) demonstrates application through an illustrative early-alert case; and (4) extends the framework to AI-enabled student services and provides practical implementation guidance.

\section{Background and Literature Review}
Learning analytics is commonly defined as the measurement, collection, analysis, and reporting of data about learners and their contexts for purposes of understanding and improving learning \cite{long2011fog}. Since its emergence, the field has pursued a strong improvement agenda, and dashboards and predictive tools are now familiar features of student success work \cite{paulsen2024dashboards}. At the same time, scholars have consistently noted that LA raises ethical questions that are educational as well as technical. Slade and Prinsloo argued that analytics depends on assumptions about what counts as normal or risky behavior \cite{slade2013ethical}, and Knox warned against treating data as neutral facts detached from institutional power \cite{knox2017data}. These concerns matter because universities do not simply observe students through analytics; they shape how students are seen and how support is organized.

Privacy, equity, and student agency are recurring themes in this literature. Jones argued that privacy in LA should be understood in relation to autonomy and informed consent \cite{jones2019consent}, and Li and colleagues showed that willingness to consent is uneven across students \cite{li2022disparities}. Sector guidance from Jisc \cite{jisc2023code} and the DELICATE checklist \cite{drachsler2016delicate} offers useful conditions for trusted analytics, though recent work argues student data rights remain underdeveloped within a broader data ecology \cite{prinsloo2024ecology}. Equity concerns are especially acute: digital traces reflect unevenly distributed social and material conditions, and student engagement is a complex, multidimensional construct that cannot be safely inferred from log data alone \cite{bergdahl2024engagement}. Capability-oriented approaches to educational justice reinforce this, shifting attention toward whether students have meaningful opportunities to succeed under differing circumstances \cite{walker2007capability}.

\section{Methodology}
This paper is a conceptual framework study based on structured literature synthesis rather than primary empirical data collection. The purpose is to develop a governance model that institutions can use when reviewing learning analytics initiatives.

The review drew on sources from six intersecting areas: learning analytics, educational data mining, student privacy, institutional governance, value-sensitive design, and educational justice. Priority was given to peer-reviewed journal articles, widely cited foundational work, and policy or legal guidance directly relevant to higher education.

The analysis proceeded in three steps. First, the literature was reviewed for recurring problem areas such as surveillance, opacity, bias, limited consent, weak accountability, and uncertain educational benefit \cite{slade2013ethical,knox2017data,jones2019consent,li2022disparities}. Second, these concerns were compared with the expectations established by legal and policy sources, including FERPA, GDPR, Jisc guidance, and the DELICATE checklist \cite{ferpa,eu2016gdpr,jisc2023code,drachsler2016delicate}. Third, themes that appeared consistently across both scholarly and governance sources were clustered into six governance dimensions. Each dimension maps to a distinct body of literature: legality to privacy and compliance scholarship; fairness to equity and capability research; student control to consent and autonomy literature; accountability to governance and policy sources; educational justification to learning analytics outcomes research; and design-stage ethics to value-sensitive design. These six dimensions became the six pillars of LEAGUE.

\section{Results}
The synthesis revealed a consistent gap: existing scholarship and governance guidance address individual dimensions of ethical concern but rarely in an integrated way. Legal frameworks address compliance; equity literature addresses fairness; consent research addresses agency; governance scholarship addresses accountability. No single institutional model connects them. LEAGUE is proposed to fill that gap by organizing these dimensions into one structured review process.

Table~\ref{tab:comparison} makes this gap explicit by comparing the ethical dimensions addressed by widely used legal and sector guidance with those addressed by LEAGUE. Each existing instrument covers part of the problem; none integrates all six dimensions into a single institutional review.

\begin{table}[htbp]
\centering
\caption{Coverage of ethical dimensions across existing guidance and the proposed LEAGUE framework (\checkmark~= addressed, P~= partial, ---~= not directly addressed).}
\label{tab:comparison}
\small
\begin{tabular}{@{}lccccccc@{}}
\toprule
\textbf{Framework} & \textbf{Law} & \textbf{Equity} & \textbf{Agency} & \textbf{Gov.} & \textbf{Utility} & \textbf{Design} & \textbf{Compreh.} \\
\midrule
FERPA \cite{ferpa} & \checkmark & --- & --- & --- & --- & --- & No \\
GDPR \cite{eu2016gdpr} & \checkmark & --- & P & \checkmark & --- & --- & No \\
DELICATE \cite{drachsler2016delicate} & \checkmark & --- & P & P & --- & \checkmark & No \\
Jisc Code \cite{jisc2023code} & \checkmark & \checkmark & P & \checkmark & P & --- & No \\
\textbf{LEAGUE} & \checkmark & \checkmark & \checkmark & \checkmark & \checkmark & \checkmark & \textbf{Yes} \\
\bottomrule
\end{tabular}
\end{table}

\subsection{The LEAGUE Framework}

\subsubsection{Lawfulness}
Lawfulness asks whether the collection, use, sharing, retention, and disclosure of student data are aligned with applicable law, policy, and contractual obligations. In higher education, this includes FERPA, GDPR where relevant, institutional privacy policy, records management rules, and vendor agreements \cite{ferpa,eu2016gdpr}. A lawful review should clarify what data are being used, for what purpose, who can access them, and how long they will be retained.

\subsubsection{Equity}
Equity asks whether the system may unfairly disadvantage some students or misread behavior shaped by context. Digital traces reflect unevenly distributed social and material conditions, and engagement inferred from log data must be treated with caution \cite{bergdahl2024engagement}. An equitable review therefore asks not only whether a technical audit finds statistical imbalance, but whether staff are trained to interpret alerts cautiously and whether interventions consider student context and meaningful opportunity to succeed \cite{walker2007capability}.

\subsubsection{Agency}
Agency concerns whether students understand how data about them are used and whether they have meaningful ways to respond, including clear notice, intelligible explanations, access to their own information, and a route for contesting inaccurate interpretations \cite{jones2019consent,li2022disparities}. A strong agency practice explains in plain language what indicators are monitored, why they matter, and how students can seek clarification.

\subsubsection{Governance}
Governance refers to the institutional structures that allocate responsibility for reviewing, approving, monitoring, and revising analytics systems. Ethical concerns often arise from unclear roles, fragmented oversight, or informal adoption rather than from the technology itself. Governance hence requires documented review processes, cross-functional oversight, role clarity, and periodic reassessment \cite{francis2023privacy}. Because predictive models and the data that feed them can exhibit non-stationarity and drift over time, periodic reassessment should revisit model performance rather than treat initial approval as final \cite{varadaraju2026pharm,varadaraju2026datgr}. A university cannot outsource responsibility for the ethical use of student data.

\subsubsection{Utility}
Utility asks whether the analytics initiative provides a clear educational purpose that justifies the data being used. Not every available data stream is educationally valuable; if a metric does not improve support, teaching, or advising, collecting it simply because it is available is difficult to justify. Systems that report LMS time-on-task as engagement, for example, have limited value if staff cannot interpret or act on that metric consistently \cite{paulsen2024dashboards}.

\subsubsection{Ethics by Design}
Ethics by Design asks if ethical questions were considered before deployment rather than only after concerns emerged, drawing on value-sensitive design \cite{friedman2019vsd}. In practice, this means testing assumptions early, involving stakeholders, reviewing message language, and building human judgment into intervention workflows. Recent work on co-designing AI-powered learning analytics shows the value of including students and teachers directly in this process \cite{alfredo2025codesign}.

\subsection{Using LEAGUE in Practice}
LEAGUE functions as an institutional review workflow applicable when a university adopts a new dashboard, expands LMS monitoring, introduces predictive advising, or revises an early alert system.

\begin{figure}[htbp]
\centering
\includegraphics[width=0.72\textwidth]{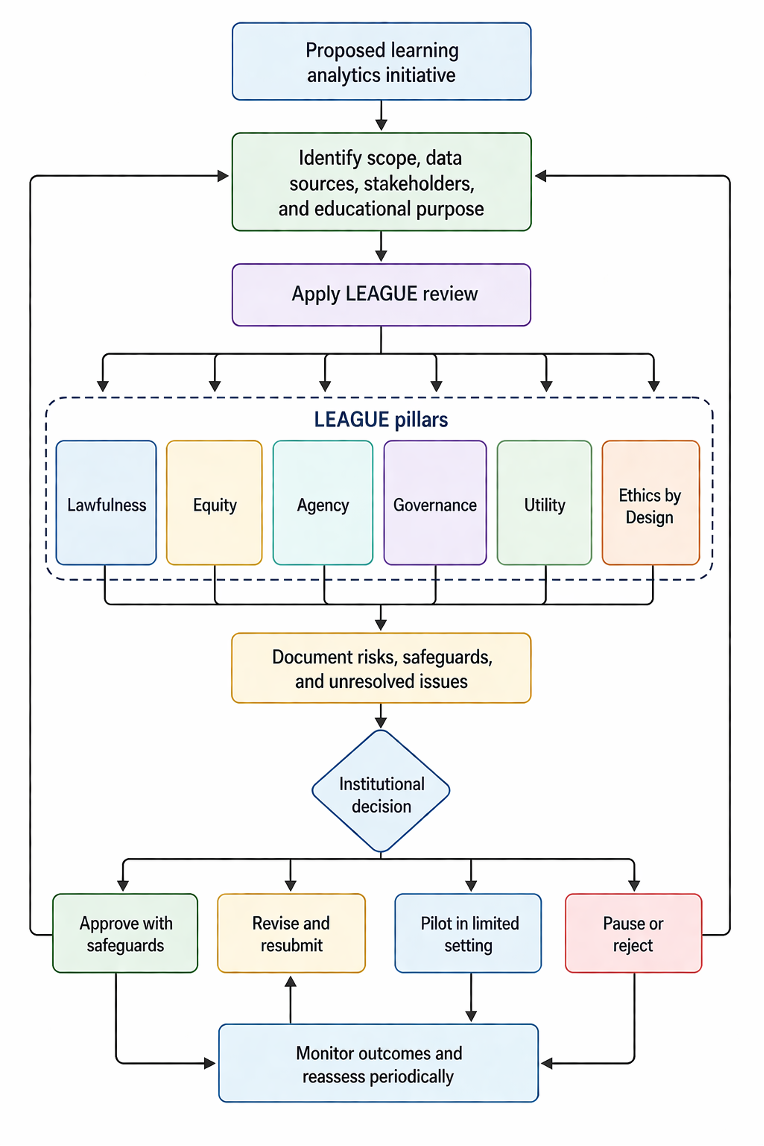}
\caption{LEAGUE review workflow for evaluating a learning analytics initiative.}
\label{fig:workflow}
\end{figure}

The workflow involves identifying the initiative, mapping data sources and stakeholders, reviewing through each LEAGUE pillar, and documenting risks and safeguards. The institution then decides whether to approve, revise, pilot, or pause the initiative, and reassesses after implementation. Fig.~\ref{fig:workflow} presents this workflow. It is designed to help committees and institutional teams use the framework in a practical, repeatable way. Table~\ref{tab:league} summarizes the guiding question, practical action, and typical evidence associated with each pillar.

\begin{table}[htbp]
\centering
\caption{Applying LEAGUE in practice.}
\label{tab:league}
\small
\renewcommand{\arraystretch}{1.3}
\begin{tabularx}{\textwidth}{@{}l X X X@{}}
\toprule
\textbf{Pillar} & \textbf{Guiding question} & \textbf{Practical action} & \textbf{Typical evidence} \\
\midrule
Lawfulness & What data are used, under what authority, and with what limits? & Map data flows, review policy basis, confirm vendor restrictions, define retention rules & Data map, privacy notice, contract clauses, retention schedule \\
Equity & Could the system misread context or disadvantage some students? & Review indicators for bias, test for uneven impact, train staff for contextual interpretation & Bias review memo, subgroup analysis, staff guidance \\
Agency & Do students understand how their data are being used? & Provide plain-language notice, allow access and clarification, create a review pathway & Student notice, FAQ, access process, case review procedure \\
Governance & Who is accountable for approval and oversight? & Assign decision-makers, establish cross-functional review, schedule reassessment & Committee charter, meeting records, review calendar \\
Utility & What educational benefit justifies the system? & Define intended outcomes, compare alternatives, evaluate actual benefit & Success metrics, evaluation report, intervention outcomes \\
Ethics by Design & Were ethical risks addressed before deployment? & Conduct impact assessment, involve stakeholders, review messages and workflows & Impact assessment, pilot feedback, design revisions \\
\bottomrule
\end{tabularx}
\end{table}

\subsection{Illustrative Case Application}
To demonstrate the operational value of LEAGUE, consider a common university scenario. An institution introduces an early alert system for first-year gateway courses. Each week, the system combines data from the LMS, attendance records, and the gradebook. Students receive a risk score from 0 to 100 based on four indicators: no LMS login for seven days, two missing assignments, attendance below 75 percent, and a current course average below 70 percent. Any student with a score of 60 or higher is added to an advisor dashboard, and the advisor is expected to send an outreach message within 48 hours.

Under Equity, the institution should ask whether the indicators are fair proxies for academic risk. A student may miss online activity because they are studying offline, commuting long hours, sharing devices, or navigating unstable connectivity. If the model treats visible online activity as the default norm, some students may be flagged more often for reasons unrelated to their actual academic commitment.

Under Governance, the institution should ask who approved the score, who can change the threshold, who reviews the model's performance, and who is responsible if the system misclassifies students. A threshold change from 60 to 40, for example, could drastically increase the number of flagged students and reshape advisor workload. That should not occur informally.

The same case also clarifies the remaining pillars. Lawfulness asks whether the combined use of LMS, attendance, and grade data is clearly authorized and whether the resulting score is treated as part of the student's educational record. Agency asks whether students know such a score is being generated and whether they can understand or question the basis of an alert. Utility asks whether the system actually improves support rather than simply producing more flags. Ethics by Design asks whether message wording, score visibility, and staff training were considered before rollout.

LEAGUE does not force a single answer. It structures the conversation so that institutional decisions are made deliberately rather than by default.

\subsection{Institutional Implementation Checklist}
The framework can be distilled into a short checklist that committees, chief information officers, institutional research offices, provosts, and ethics review boards can apply before approving or renewing a learning analytics system:

\begin{itemize}
\item[$\square$] \textbf{Lawfulness:} Is there a clear legal and contractual basis for the data used, with a defined purpose and retention period?
\item[$\square$] \textbf{Equity:} Has subgroup bias been evaluated, and are staff trained to interpret alerts in context?
\item[$\square$] \textbf{Agency:} Can students understand how their data are used and contest inaccurate interpretations?
\item[$\square$] \textbf{Governance:} Is there clear ownership for approval, monitoring, and periodic reassessment?
\item[$\square$] \textbf{Utility:} Does the system demonstrably improve teaching, advising, or student support?
\item[$\square$] \textbf{Ethics by Design:} Were ethical risks considered before deployment, rather than only after concerns emerged?
\end{itemize}

\section{Discussion}
The LEAGUE framework contributes to the learning analytics literature by bringing several strands of concern into one institutional model. It does not replace legal review, technical audit, or local policy development. Instead, it connects them: privacy and fairness are considered alongside student agency and educational context, while accountability is treated as an organizational responsibility and educational benefit is asked for explicitly rather than assumed.

This integrative feature is especially useful for practice. Different institutional actors often focus on different parts of the problem: legal counsel on compliance, information technology on infrastructure, student success units on intervention, and faculty on teaching implications. LEAGUE offers a common language through which these groups can review the same initiative together. It also responds to recent case-based work showing that privacy in student success systems is often handled unevenly in practice, even when institutions see themselves as compliant \cite{francis2023privacy}. Practical experience deploying context-aware institutional AI, such as prompt-engineered course-discovery systems that rely on feedback loops and guardrails, reinforces that such safeguards must be operationalized rather than assumed \cite{vijayakumar2026coursediscovery}.

Because responsibility for these dimensions is distributed, LEAGUE also helps clarify ownership. Table~\ref{tab:stakeholders} offers an illustrative mapping of pillars to the stakeholders typically best positioned to lead each review; in practice most pillars are shared, and the mapping is intended to start rather than settle the conversation about accountability.

\begin{table}[htbp]
\centering
\caption{Illustrative mapping of LEAGUE pillars to institutional stakeholders.}
\label{tab:stakeholders}
\small
\begin{tabular}{@{}ll@{}}
\toprule
\textbf{Stakeholder} & \textbf{Primary LEAGUE pillar} \\
\midrule
Registrar & Lawfulness \\
Chief Information Officer & Governance \\
Faculty & Utility \\
Institutional Research & Equity \\
Students & Agency \\
Ethics / IRB committee & Ethics by Design \\
\bottomrule
\end{tabular}
\end{table}

\subsection{LEAGUE and AI-Enabled Student Services}
As universities increasingly deploy LLM-powered advising, AI chatbots, agentic systems, and predictive copilots, governance challenges extend beyond dashboards to conversational systems, automated recommendations, and generative AI. These systems raise the same six questions in sharper form: the legal basis for training and prompt data, equity across student populations, student awareness of automated interactions, ownership of model behavior, genuine educational value, and design-stage ethical review. LEAGUE provides a governance structure that applies to both conventional learning analytics and AI-enabled student services, offering continuity as institutions move from dashboards to generative and agentic tools \cite{vijayakumar2026coursediscovery}.

\subsection{A Governance Maturity Perspective}
LEAGUE can also be read as a maturity progression. Institutions frequently begin at a compliance-only posture and can advance toward continuous, AI-ready governance, as summarized in Table~\ref{tab:maturity}.

\begin{table}[htbp]
\centering
\caption{Governance maturity levels for learning analytics.}
\label{tab:maturity}
\small
\renewcommand{\arraystretch}{1.2}
\begin{tabularx}{\textwidth}{@{}lX@{}}
\toprule
\textbf{Level} & \textbf{Description} \\
\midrule
1. Compliance only & Review is limited to legal permissibility (e.g., FERPA and GDPR). \\
2. Documented governance & Roles, approvals, and data practices are recorded and repeatable. \\
3. Ethical governance & Equity, agency, and educational utility are reviewed alongside compliance. \\
4. Continuous monitoring & Systems are reassessed on a schedule, including model performance and drift. \\
5. AI-ready governance & The same review extends to LLM, agentic, and generative student-facing systems. \\
\bottomrule
\end{tabularx}
\end{table}

\subsection{Limitations and Future Work}
The framework has clear limits. It is a conceptual model rather than an empirically validated instrument, and it is not a substitute for technical fairness testing or legal counsel. Several directions would strengthen it: expert review to refine the pillars and their guiding questions; application and case studies across diverse institutions to test practical fit; evaluation of international applicability beyond the FERPA and GDPR contexts emphasized here; and development of quantitative assessment rubrics that would let institutions score and track governance maturity over time. A natural progression is to move from this conceptual framework toward a validated instrument suitable for empirical study.

\section{Conclusions}
Higher education institutions are under growing pressure to use student data strategically, but the ethical quality of governance matters as much as the predictive power of the tools being used. Compliance alone is not enough: ethical governance of learning analytics also requires attention to equity, student agency, institutional accountability, educational usefulness, and design-stage reflection. The LEAGUE framework was proposed to help institutions address these concerns in a structured way.

Ultimately, the question is not only whether a university can use student data in a particular way, but whether it can do so fairly, transparently, and in support of meaningful educational purposes. If learning analytics is to fulfill its promise, it must be governed in ways that students can trust.

\section*{Acknowledgements}
The authors thank Jacqueline M. Ring, Vice Chancellor and Chief Institutional Research Officer at Rowan University, for her support and guidance throughout this work.

\bibliographystyle{IEEEtran}
\bibliography{refs}

@article{long2011fog,
	author  = {Long, P. and Siemens, G.},
	title   = {Penetrating the Fog: Analytics in Learning and Education},
	journal = {EDUCAUSE Review},
	volume  = {46},
	number  = {5},
	year    = {2011},
	url     = {https://er.educause.edu/articles/2011/9/penetrating-the-fog-analytics-in-learning-and-education}
}

@article{macfadyen2010mining,
	author  = {Macfadyen, L. P. and Dawson, S.},
	title   = {Mining LMS Data to Develop an Early Warning System for Educators: A Proof of Concept},
	journal = {Computers \& Education},
	volume  = {54},
	number  = {2},
	pages   = {588--599},
	year    = {2010},
	doi     = {10.1016/j.compedu.2009.09.008}
}

@article{slade2013ethical,
	author  = {Slade, S. and Prinsloo, P.},
	title   = {Learning Analytics: Ethical Issues and Dilemmas},
	journal = {American Behavioral Scientist},
	volume  = {57},
	number  = {10},
	pages   = {1510--1529},
	year    = {2013},
	doi     = {10.1177/0002764213479366}
}

@article{knox2017data,
	author  = {Knox, J.},
	title   = {Data Power in Education: Exploring Critical Awareness with the Learning Analytics Report Card},
	journal = {Television and New Media},
	volume  = {18},
	number  = {8},
	pages   = {734--752},
	year    = {2017},
	doi     = {10.1177/1527476417690029}
}

@misc{ferpa,
	author       = {{U.S. Department of Education, Student Privacy Policy Office}},
	title        = {Family Educational Rights and Privacy Act (FERPA)},
	year         = {2026},
	note         = {Accessed May 2026},
	howpublished = {\url{https://studentprivacy.ed.gov/content/family-educational-rights-and-privacy-act}}
}

@misc{eu2016gdpr,
	author       = {{European Union}},
	title        = {Regulation (EU) 2016/679 of the European Parliament and of the Council of 27 April 2016 (General Data Protection Regulation)},
	year         = {2016},
	note         = {Accessed May 2026},
	howpublished = {\url{https://eur-lex.europa.eu/legal-content/EN/TXT/?uri=CELEX:32016R0679}}
}

@misc{jisc2023code,
	author       = {{Jisc}},
	title        = {Code of Practice for Learning Analytics},
	year         = {2023},
	note         = {Accessed May 2026},
	howpublished = {\url{https://www.jisc.ac.uk/guides/code-of-practice-for-learning-analytics/}}
}

@inproceedings{drachsler2016delicate,
	author    = {Drachsler, H. and Greller, W.},
	title     = {Privacy and Analytics -- It's a DELICATE Issue: A Checklist for Trusted Learning Analytics},
	booktitle = {Proceedings of the Sixth International Conference on Learning Analytics and Knowledge},
	pages     = {89--98},
	year      = {2016},
	doi       = {10.1145/2883851.2883893}
}

@article{paulsen2024dashboards,
	author  = {Paulsen, L. and Lindsay, E.},
	title   = {Learning Analytics Dashboards Are Increasingly Becoming About Learning and Not Just Analytics -- A Systematic Review},
	journal = {Education and Information Technologies},
	volume  = {29},
	number  = {11},
	pages   = {14279--14308},
	year    = {2024},
	doi     = {10.1007/s10639-023-12401-4}
}

@article{jones2019consent,
	author  = {Jones, K. M. L.},
	title   = {Learning Analytics and Higher Education: A Proposed Model for Establishing Informed Consent Mechanisms to Promote Student Privacy and Autonomy},
	journal = {International Journal of Educational Technology in Higher Education},
	volume  = {16},
	pages   = {24},
	year    = {2019},
	doi     = {10.1186/s41239-019-0155-0}
}

@article{li2022disparities,
	author  = {Li, W. and Sun, K. and Schaub, F. and Brooks, C.},
	title   = {Disparities in Students' Propensity to Consent to Learning Analytics},
	journal = {International Journal of Artificial Intelligence in Education},
	volume  = {32},
	number  = {3},
	pages   = {564--608},
	year    = {2022},
	doi     = {10.1007/s40593-021-00254-2}
}

@article{prinsloo2024ecology,
	author  = {Prinsloo, P. and Khalil, M. and Slade, S.},
	title   = {Learning Analytics as Data Ecology: A Tentative Proposal},
	journal = {Journal of Computing in Higher Education},
	volume  = {36},
	pages   = {154--182},
	year    = {2024},
	doi     = {10.1007/s12528-023-09355-4}
}

@article{bergdahl2024engagement,
	author  = {Bergdahl, N. and Bond, M. and Sj{\"o}berg, J. and Dougherty, M. and others},
	title   = {Unpacking Student Engagement in Higher Education Learning Analytics: A Systematic Review},
	journal = {International Journal of Educational Technology in Higher Education},
	volume  = {21},
	pages   = {63},
	year    = {2024},
	doi     = {10.1186/s41239-024-00493-y}
}

@book{walker2007capability,
	editor    = {Walker, M. and Unterhalter, E.},
	title     = {Amartya Sen's Capability Approach and Social Justice in Education},
	publisher = {Palgrave Macmillan},
	address   = {New York},
	year      = {2007}
}

@article{alfredo2025codesign,
	author  = {Alfredo, R. and Milesi, M. and Echeverria, V. and others},
	title   = {Co-designing AI-Powered Learning Analytics: Bringing Students and Teachers Together},
	journal = {International Journal of Educational Technology in Higher Education},
	volume  = {22},
	pages   = {78},
	year    = {2025},
	doi     = {10.1186/s41239-025-00572-8}
}

@book{friedman2019vsd,
	author    = {Friedman, B. and Hendry, D. G.},
	title     = {Value Sensitive Design: Shaping Technology with Moral Imagination},
	publisher = {MIT Press},
	address   = {Cambridge, MA},
	year      = {2019}
}

@article{francis2023privacy,
	author  = {Francis, M. and Avoseh, M. B. M. and Card, K. and Newland, L. and others},
	title   = {Student Privacy and Learning Analytics: Investigating the Application of Privacy Within a Student Success Information System in Higher Education},
	journal = {Journal of Learning Analytics},
	volume  = {10},
	number  = {3},
	pages   = {102--114},
	year    = {2023},
	doi     = {10.18608/jla.2023.7975}
}

@inproceedings{varadaraju2026pharm,
	author    = {Varadaraju, Sahana and Vijayakumar, Bharathwaj},
	title     = {Pharmacological Non-Stationarity in Human--AI Systems: A Framework for Medication-Aware Adaptive Decision Support},
	booktitle = {2026 IEEE International Conference on Human-Machine Systems (ICHMS)},
	year      = {2026},
	pages     = {616--621},
	doi       = {10.1109/ICHMS69701.2026.11602193}
}

@inproceedings{vijayakumar2026coursediscovery,
	author    = {Vijayakumar, Bharathwaj and Varadaraju, Sahana K. and Balaji Baskaran, Lathish and Bouaynaya, Nidhal C. and Alapati, Samyukta},
	title     = {AI-Powered Course Discovery: Leveraging Prompt Engineering, Feedback Loops, and Guardrails for Institutional Context},
	booktitle = {2026 IEEE Global Engineering Education Conference (EDUCON)},
	year      = {2026},
	pages     = {1--5},
	doi       = {10.1109/EDUCON67543.2026.11574417}
}

@inproceedings{varadaraju2026datgr,
	author    = {Vijayakumar, Bharathwaj and Varadaraju, Sahana K.},
	title     = {Drift-Aware Temporal Graph Rewiring (DATGR) for Adaptive Semantic Modeling in Biomedical Text},
	booktitle = {Proceedings of the 2026 IEEE Conference on Artificial Intelligence (CAI)},
	year      = {2026},
	pages     = {193--198},
	doi       = {10.1109/CAI68641.2026.11536228}
}

\end{document}